\documentstyle[sprocl]{article}

\def\Journal#1#2#3#4{{#1} {\bf #2}, #3 (#4)}

\def\PLB{{\em Phys. Lett.}  B}

\def\ra{\rightarrow}

\def\be{\begin{equation}}
\def\ee{\end{equation}}
\def\bea{\begin{eqnarray}}
\def\eea{\end{eqnarray}}

\begin{document}

\title{NONLEPTONIC DECAYS: AMPLITUDE ANALYSIS AND SUPERMULTIPLET
SCHEMES}

\author{R. Delbourgo and Dongsheng Liu}

\address{School of Physics, University of Tasmania,\\
 GPO Box 252-21, Hobart, AUSTRALIA 7001 \\
 E-mail: Bob.Delbourgo@utas.edu.au, D.Liu@utas.edu.au} 




\maketitle\abstracts{We give best covariant amplitude decompositions
 for two-body decay processes involving ground state hadrons ($0^-,
 1^-, \frac{1}{2}^+, \frac{3}{2}^+$) and show how these are simply
 related to helicity amplitudes. After discussing how electromagnetic
 interactions are incorporated via meson dominance in a relativistic
 supermultiplet scheme, we extend the analysis to weak flavour
 changing nonleptonic processes. Such weak interactions are described
 by three generic amplitudes, which we estimate according to the rules
 of calculation within covariant SU(2N$_f$).}

\section{Introduction}
The advent of heavy meson factories and the discovery of new heavy
baryon states has led to a veritable explosion of new data on the 
decays of such hadrons. A high fraction of these decays occur through
nonleptonic channels and it becomes an urgent matter to gain a full
understanding\footnote{Semileptonic amplitudes are pretty well
understood in terms of the current-current picture, using a number of
`decay constants' and `form factors' which parametrise them.}
of how these occur. In the first place we should try to arrive at a 
comprehensive and reasonably accurate description for such
nonleptonic amplitudes, since a gross picture using the Fermi constant
$G_F$, the CKM elements $V_{UD}$ and appropriate phase space factors,
is already able to predict the widths within a factor of about two.
Several~\cite{Revs} reviews of this subject have appeared in which the
virtues and failings of different descriptions of nonleptonic models
are discussed.

In this paper we shall exhibit a scheme for calculating the weak 
flavour-changing amplitudes which is founded on a relativistic
supermultiplet scheme that already agrees with measured strong and 
electromagnetic 3-particle amplitudes~\cite{RDDL1} within 10\% or better. 
We are hopeful that it will account for experimentally measured 
nonleptonic decays to roughly the same accuracy, since it allows 
for mixing between weak states (which are normally rotated away into 
the strong interactions by a physical state mass diagonalization) in 
addition to the usual $W$-exchange contributions.

In the next section we summarise how the weak amplitudes are best
expressed \cite{RDDL2}, because of their simple connection with
helicity amplitudes. Following this analysis we recall the main features
of relativistic SU(2N$_f$) and the rules for calculating strong
interaction amplitudes. After a digression into how meson dominance 
is implemented in such a supermultiplet framework, we turn to weak 
interactions. We show how all processes to order $G_F$ naturally
subdivide into three types and we describe how these three amplitudes
may be estimated. In future research we shall apply these evaluative
rules to physical decays; this is a major task as there is such a
multitude of nonleptonic processes and our ambition is to predict 
them {\em all} to the same accuracy as strong and electromagnetic 
(em) processes.

\section{Weak Amplitude Analysis}
\subsection{Counting the Amplitudes}
Let us focus on two-body decays, since these must be properly
comprehended before many-body decays. Consider the process\footnote{
The choice of momentum is for ease of crossing: one simply interchanges
particle labels, includes the conjugation factor $(-1)^{2j}$ and 
modifies the spin averaging and mass factors to derive the crossed
decay rate.} $-p_1j_1\!\rightarrow\!p_2j_2 + p_3j_3$, 
associated with elements,
$\langle p_2\lambda_2,p_3\lambda_3|S|-p_1\lambda_1 \rangle =
  (2\pi)^4\sum_{\{\lambda\}} \delta^4(p_1+p_2+p_3) M_{\{\lambda\}},$
and leading to the decay rate
$$\Gamma=\frac{\sum_\lambda\Delta|M_{\{\lambda\}}|^2}{16\pi m_1^3(2j_1+1)};
\quad \Delta^2=m_1^4+m_2^4+m_3^4-2m_1^2m_2^2-2m_2^2m_3^2-2m_3^2m_1^2.$$
Now the number of independent couplings of these particles equals the
number of ways $N$ in which the three spins can be coupled to produce 
an integer angular momentum. If the $j_i$ form a Euclidean triangle,
$$N=j_1(1-j_1)+j_2(1-j_2)+j_3(1-j_3)+2(j_1j_2+j_2j_3+j_3j_1)+1.$$
Otherwise arrange them in decreasing order: $j_a\!\geq\!j_b\!\geq\!j_c$,
with $j_a\!\geq\!j_b\!+\!j_c$; then $N=(2j_b+1)(2j_c+1).$ [The same 
result can be deduced via the standard $L-S$ coupling method or via 
the helicity formalism.]  No assumption about parity conservation has 
been made since we shall largely concentrate on weak amplitudes; 
nor have we supposed the couplings to be real, because they are 
surely not in the decay region as a result of strong final state 
interactions. The values of the couplings will depend on the masses 
$m_i$ and the quantum numbers of the particles. Generally, we should 
expect that the larger the difference in the spin values (excitation 
numbers) and quantum numbers (flavour values) the smaller the couplings 
will be because of decreased overlap between the particle
`wave-functions'. 

\subsection{Covariant Decompositions}
In Lorentz covariant descriptions regards the amplitudes 
as arising from an effective three-particle Lagrangian
\be
{\cal L} = \sum_{I=1}^N g_I^{j_1j_2j_3}
                  \phi^{j_1}(-p_1)\phi^{*j_2}(p_2)\phi^{*j_3}(p_3),
\ee
where $\phi^j(p)$ stands for a free-field solution, namely
$1,u^\lambda(p),\epsilon_\mu^\lambda(p),u_\mu^\lambda(p)\ldots$ for
particles of spin $0,\frac{1}{2},1,\frac{3}{2}\ldots$. We give two 
examples of how the couplings are best defined within the covariant 
formalism, as they illustrate the basic simplicity of the resulting
expressions. A complete set of decompositions is provided in Ref 3.
\vspace{.2cm}

\noindent
\underline{$1/2 \rightarrow 1/2 + 1$}\,\,\,(e.g. $\Sigma^+\rightarrow 
 p\gamma,\quad \Lambda_b\rightarrow\Lambda\psi$).

\noindent
The couplings are most neatly expressed in the Sachs form,
\be
{\cal L}=\epsilon_3^{*\nu}\bar{u}_2[(p_2-p_1)_\nu
         (f_E+g_E\gamma_5)/2 + \epsilon_{\nu\rho\sigma\tau}
         p_1^\rho p_2^\sigma \gamma^\tau(f_M\gamma_5+g_M)]u_1,
\ee
where the subscripts $M,E$ refer to the `electric' and `magnetic'
parts of the vector interaction. It is not recommended to use the
more traditional decompositions $\gamma_\nu,\gamma_\nu\gamma_5,
\sigma_{\nu\rho}p_3^\rho, \sigma_{\nu\rho}\gamma_5p_3^\rho$, because
they lead to complicated cross-terms in decay rates and the like;
also they are more distantly related to helicity amplitudes.

However, with decomposition (2), the ensuing decay rate looks neat:
\bea
\Gamma =& \frac{\Delta^3}{32\pi m_1^3}[&((m_1m_2-p_1\cdot p_2)
          (|f_E|^2/m_3^2 + 2|f_M|^2)  \nonumber        \\
        & & -(m_1m_2+p_1\cdot p_2)(|g_E|^2/m_3^2 + 2|g_M|^2)].
\eea
If the final vector particle is a photon, gauge invariance ensures
that $f_E,\! g_E \rightarrow 0$ (remember that $m_1\neq m_2$),
whereupon the rate simplifies further to
\be
\Gamma \rightarrow \frac{(m_1^2-m_2^2)^3}{16\pi m_1^3}
       \left[(m_1+m_2)^2|f_M|^2 + (m_1-m_2)^2|g_M|^2\right],
\ee
and one of $f_M$ or $g_M$ must be discarded because of em parity
conservation.
\vspace{.2cm}

\noindent
\underline{$1 \rightarrow 1 + 1$}\,\,\,(e.g. $\Psi\rightarrow 
 K^*\bar{K}^*,\,\, \gamma\gamma$).

\noindent
As far as we know, the tidiest decomposition has ${\cal L}=
\epsilon_2^{*\mu}\epsilon_3^{*\nu}{\cal M}_{\lambda\mu\nu}
\epsilon_1^\lambda$ with
\bea
{\cal M}_{\lambda\mu\nu}&=&(p_3-p_2)_\lambda
     [g_{1T}(p_2\cdot p_3\eta_{\mu\nu}-p_{3\mu}p_{2\nu}) + g_{1M}
         \epsilon_{\mu\nu\rho\sigma} p_2^\rho p_3^\sigma]/2\nonumber\\
& & +{\rm cyclic} + g_L(p_3-p_2)_\lambda(p_1-p_3)_\mu(p_2-p_1)_\nu/8.
\eea
The expression for the decay rate substantiates our claim:
\bea
\Gamma &=& \frac{\Delta^3}{192\pi m_1^3}\left[\frac{\Delta^4}
          {16m_1^2m_2^2m_3^2}|g_L|^2 + \frac{1}{2}\Re(g_L^*
              \sum_i \frac{g_{iT}}{m_i^2}) + \right.
          \nonumber \\
  & & \left. \sum_{i\neq j\neq k}\left (2\Re(m_i^2g_{jT}g_{kT}^*) +
  \frac{(\Delta^2 +m_j^2m_k^2)|g_{iT}|^2 + \Delta^2|g_{iM}|^2}{2m_i^2}
         \right) \right].
\eea
Once again, simplifications arise when one vector meson is a photon,
say particle 2; in that case, $g_L, g_{2M}, g_{2T} \rightarrow 0$, and
\be
 \Gamma \rightarrow \frac{\Delta^5}{384\pi m_1^3}\left[
        \frac{|g_{1T}|^2 + |g_{1M}|^2}{m_1^2} +
        \frac{|g_{3T}|^2 + |g_{3M}|^2}{m_3^2} \right].
\ee

These examples are but two cases of the entire set involving ground
state hadrons, which are listed in Ref 3; they nicely illustrate 
the advantages of an elegant covariant decomposition. 

\subsection {Helicity Amplitudes}
As further confirmation of the `best coupling' covariants we may work 
out\footnote{Recall that parity conservation, encapsulated by 
$M_{\lambda_2,\lambda_3} = \eta_1\eta_2\eta_3(-1)^{j_2+j_3-j_1}
M_{-\lambda_2,-\lambda_3}$, will roughly halve the number of couplings.
} the helicity amplitudes $M_{\lambda_2,\lambda_3}$. The idea is that
they should be very simple linear combinations of the chosen couplings
$g_I^{j_1j_2j_3}$; and indeed they are. With our two examples one finds
\vspace{.2cm}

\noindent
\underline{$1/2 \rightarrow 1/2 + 1$}

\noindent
$$M_{1/2,0}+M_{-1/2,0}=\sqrt{2(m_1m_2-p_1\cdot p_2)}\Delta f_E/m_3,$$
$$M_{1/2,0}-M_{-1/2,0}=\sqrt{2(m_1m_2+p_1\cdot p_2)}\Delta g_E/m_3,$$
$$M_{1/2,1}+M_{-1/2,1}=-2\sqrt{m_1m_2-p_1\cdot p_2}\Delta f_M,$$
$$M_{1/2,1}-M_{-1/2,1}=-2\sqrt{m_1m_2+p_1\cdot p_2}\Delta g_M.$$
Notice how the threshold factors of relative momentum turn up
automatically and how simple it is to continue from coupling $f$ to 
its parity opposite $g$.  Of course the helicity state $\lambda_3=0$ 
and amplitude must be ignored for photons.
\vspace{.2cm}

\noindent
\underline{$1 \rightarrow 1 + 1$}

\noindent
$$M_{0,0} = -\frac{1}{2}\Delta m_1m_2m_3\left(\frac{g_{1T}}{m_1^2}+
              \frac{g_{2T}}{m_2^2}+\frac{g_{3T}}{m_3^2} + 
              \frac{\Delta^2g_L}{4m_1^2m_2^2m_3^2}\right),$$
while the other well-defined parity combinations are
$$M_{1,1} + M_{-1,-1} = p_2\cdot p_3\Delta g_{1T}/m_1,\quad
  M_{1,1} - M_{-1,-1} = i\Delta^2 g_{1M}/2m_1, $$
$$M_{0,1} + M_{0,-1} = i\Delta^2 g_{2M}/2m_2,\quad
  M_{0,1} - M_{0,-1} = p_2\cdot p_3\Delta g_{2T}/m_2, $$
$$M_{1,0} + M_{-1,0} = i\Delta^2 g_{3M}/2m_3,\quad
  M_{1,0} - M_{-1,0} = -p_2\cdot p_3\Delta g_{3T}/m_3, $$
whereupon we see why $g_{2T}, g_{2M}$ and $g_L$ have to discarded
when 2 is a photon.

\section{Strong and Electromagnetic Interactions of Supermultiplets}

\subsection{Supermultiplet Wavefunctions}
We now turn to the strong hadronic interactions, which are believed
to be described by the fundamental chromodynamic Lagrangian,
\be
{\cal L} = \sum_{f,colour}\bar{\psi}_f(i\!\not\!\partial -
        g\!\not\! A.\lambda/2 - M_f)\psi_f - F_{\mu\nu}F^{\mu\nu}/4.
\ee
This possesses a heavy quark symmetry \cite{HQS} for 
$M_f\gg\Lambda_{\rm QCD}$ corresponding to `flavour-blindness' of 
quarks moving with equal velocity $v$. In fact, if one neglects
the gluon field altogether (by doing a Foldy-Wouthuysen transformation 
\cite{KT}to leading order) one finds that for hadrons bound by 
{\em constituent} quarks, the Lagrangian has \cite{DSS} a 
[U(2N$_f$)$\otimes$U(2N$_f$)]$_v$ symmetry, regardless of $m_q$.
In this traditional picture of hadrons, the baryons and mesons are
composed of quark moving in tandem with the {\em same} velocity $v$
and all gluon effects are taken into account in dressing the quarks
from `current' to `constituent' with little or no binding energy.

For instance, mesons are described by the wavefunction
$\Phi_A^B\!\equiv\!u_A(p_1)\bar{v}^B(p_2)$ which can be pictured as
two parallel quark lines with symbols $A=\alpha a, B=\beta b$ 
carrying the Dirac spinor labels $\alpha,\beta$ and flavour labels
$a,b$. Since the quarks have equal velocity $v = p/(m_1+m_2) 
\equiv p/\mu$, we may write $p_i=m_iv$, where $p$ is the total 
momentum of the meson, whereupon we see that, {\em even for unequal 
mass quarks} the meson wavefunction obeys the Bargmann-Wigner 
equations for a bispinor,
$(\not\!p-\mu)\Phi=\Phi(\not\!p+\mu)=0,\quad {\rm with} 
 \quad\Phi_A^B(p)\equiv u_A(p)\bar{v}^B(p).$
It follows that
\be
\Phi_A^B(p)=[(\not\!p+\mu)(\gamma_5\phi_{a5}^b-\gamma^\mu\phi_{a\mu}^b)
     ]_\alpha^\beta \equiv 2\mu P_{+v}(\gamma_5\phi_5-\gamma\cdot\phi),
\ee
where $P_{+v}\equiv (1+\!\not\!v)/2$. Similarly, for baryons one finds
\bea
\Psi_{(ABC)}(p) & = & [P_{+v}\gamma_\mu C]_{\alpha\beta}
               u^\mu_{(abc)\gamma} + \frac{\sqrt{2}}{3}\left(
   [P_{+v}\gamma_5C]_{\alpha\beta} u_{[ab]c\gamma}+\right.\nonumber\\
 & & \left.\qquad [P_{+v}\gamma_5C]_{\beta\gamma} u_{[bc]a\alpha} +
   [P_{+v}\gamma_5C]_{\gamma\alpha} u_{[ca]b\beta}\right).
\eea
The basic three-point vertices between ground state mesons and baryons, 
maintaining the maximal symmetry U$_w$(2N$_f$) are governed by just two
couplings\footnote{Actually the dimensionless couplings to be used are
$g=3G\Sigma/4$ and $f=F\mu\Sigma$, where $\Sigma$ is the sum of the
participating masses.}, $F$ and $G$:
\bea
{\cal L} & = & F\Phi_A^B(p_1)\left( \Phi_B^C(p_2)\Phi_C^A(p_3) +
                 \Phi_B^C(p_3)\Phi_C^A(p_2) \right)  \nonumber\\
   & &  +  G\bar{\Psi}^{ABC}(p_2)\Phi_C^D(-p_3)\Psi_{DAB}(p_1).
\eea
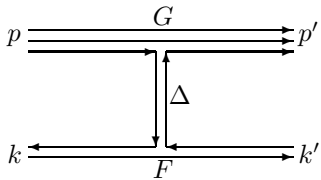
\begin{figure}
\begin{picture}(200,100)(0,-40)
\put(92,4){$p$}
\put(100,8){\vector(1,0){100}}
\put(202,4){$p'$}
\put(147,10){$G$}
\put(100,4){\vector(1,0){100}}
\put(100,0){\vector(1,0){48}}
\put(152,0){\vector(1,0){48}}
\put(148,0){\vector(0,-1){36}}
\put(152,-36){\vector(0,1){36}}
\put(148,-36){\vector(-1,0){48}}
\put(200,-36){\vector(-1,0){48}}
\put(100,-40){\vector(1,0){100}}
\put(92,-42){$k$}
\put(202,-42){$k'$}
\put(147,-48){$F$}
\put(153,-20){$\Delta$}
\end{picture}
\caption{A duality diagram or multispinor trace corresponding
to meson-baryon scattering associated with a meson pole. 
\label{fig:Figure1}}
\end{figure}

In such relativistic supermultiplet schemes, all scattering amplitudes
are obtained as tree diagrams by sewing together the multispinors and
vertices. No loop corrections are to be evaluated since those quantum 
loops really correspond to QCD gluon effects and, by assumption, 
those have already been largely taken into account into dressing the 
constituent quarks in the hadrons. For example the process $\pi(k)N(p)
\rightarrow \rho(k') N(p')$ due to $\omega$ and $\pi$ exchange is 
contained within the expression,
$$ GF\bar{\Psi}^{ABC}(p')\Psi_{ABD}(p)\Delta_{CF}^{DE}(p-p')
     \Phi_E^G(k)\Phi_G^F(-k');\quad \Delta = {\rm meson~propagator},$$
or as a duality diagram drawn in Figure 1.  The constants $F$ and $G$
are related to the $\rho N N$ and $\rho \pi\pi$ couplings, and can be
further related by invoking Sakurai's isospin universality hypothesis.
One can write a similar expression for the baryon pole term and
even add a four-point contact term to represent non-resonant 
contributions to the process. A large body of data on strong and em 
interactions (via vector dominance) can be tied within such a higher 
symmetry scheme and it is a source of wonder---especially for the
lighter hadrons where the symmetry is on shakier ground---that the 
theoretical predictions agree \cite{RDDL1} with the experimental facts 
to within 10\% and generally much better, using as inputs $g_{\pi NN}$ 
and the average supermultiplet constituent masses
$$ m_{ud}\simeq 350,\,\,m_s\simeq 450,\,\,m_c\simeq 1500,\,\,
 m_b\simeq 4700\quad ({\rm in~MeV}).$$

\subsection{Vector Dominance}
There exist a number of implementations of vector dominance. The 
version we shall adopt has a photon ($A$) vector meson ($V$) coupling
$-eA\cdot F_V(\partial^2\!)V/g_Vm_V^2$, where $F_V$ is a `form-factor' 
describing the off-shell mixing. By convention, the normalization
$F(0)=m_V^2$ ensures that the quark charges come out correctly at
soft photon momentum, provided that the vector coupling $g_V$ to the
quarks is normalized by the appropriate CG coefficient; specifically,
\be
g_\rho:g_\omega:g_\phi:g_\psi:g_\Upsilon=\sqrt{2}:3\sqrt{2}:-3:3/2:-3;
\quad g_\rho=3.54\pm .14.
\ee
One can get a good idea of how $F_V$ varies with $m_V$ by examining the
leptonic decay widths of the vector mesons,
$\Gamma_{V\ra l\bar{l}} \simeq \frac{m_V}{12\pi}
  \left(\frac{e^2F_V}{g_V}\right)^2.$
The experimental results strongly suggest that, once the CG factor is
subsumed in $g_V$ as above, the widths vary little. Thus one is led to
{\em assume} that $F_V^2 \propto 1/m_V$, although more sophisticated
extrapolations, based on the renormalization group, exist. Thus
we take $\langle 0|j_\mu^{em}|V\rangle = em_V^2F_V\epsilon_\mu/g_V\equiv 
em_Vf_V\epsilon_\mu$, where the `meson decay constant' $f_V$ equals 
$m_VF_V/g_V \simeq \sqrt{\Lambda m_V}/g_\rho$ with $\Lambda = m_\rho.$  
A comparison between these crude theoretical predictions and the 
experimental values of $f_V$, derived from leptonic decay widths, is
provided in Table 1.
\begin{table}
\caption{Experimental values of vector meson decay constants $f_V$,
 in MeV as determined from leptonic decay widths, via  $(e^2 f_V)^2= 
 12\pi m_V\Gamma_V$, compared with the theoretical predictions,
 $(g_Vf_V)^2 = \Lambda m_V$. Experimental errors hover around 5\%.
 Heavier mesons values are inferred as there is no {\em direct}
 experimental data for them.}
\vspace{0.2cm}
\begin{center}
\begin{tabular}{|l|c|c|c|c|c|c|c|c|c|} 
\hline
Vector Meson & $\rho$ & $\omega$ & $\phi$ & $\psi$ & $\Upsilon$ &
 $K^*$ & $D^*$ & $D_s^*$ & $B^*$\\ 
\hline
$|f_V|$ (expt.) & 154 & 47 & 80 & 274 & 240 & ? & ?  & ? & ? \\
\hline
$|f_V|$ (theory) & 154 & 51 & 83 & 281 & 252 & 232 & 347 & 355 & 565 \\
\hline
\end{tabular}
\end{center}
\end{table}
In supermultiplet terms one is led to introduce the em field multispinor
${\cal A}_B^C = e(\gamma\cdot A)_\beta^\gamma Q_a^b$ ($Q$ is the flavour 
charge matrix) and the contact interaction
$${\cal L} = {\rm Tr}[{\cal A} \sqrt{\Lambda m_V}\Phi/g_V]/4
 = em_Vf_V A\cdot\phi, $$
as required. Note that as before, we do {\em not} rotate the em
and vector field in order to diagonalize the masses, according to
$V'\!=\!(V + eA/g_V)/\sqrt{1+e^2/g_V^2}$ and
 $A'\!= (A - eV/g_V)/\sqrt{1+e^2/g_V^2}.$
This would otherwise muddy the leptonic interaction,
$$ {\cal L}_l=e\bar{l}\!\not\!A l=e\bar{l}\!(\not\!A'+e\not\!V'/g_V)l
                /\sqrt{1+e^2/g_V^2}, $$
by introducing an intrinsic interaction of the lepton with the strong 
vector meson, and complicating the picture.  It is much easier to use
the original $A$ and $V$ fields, which mix through quark interactions,
and thereby allow a $A$-$V$ contact term as is done in most 
interpretations of vector dominance. We will adopt exactly the same 
approach for weak interactions.

\section{Weak Flavour-changing Nonleptonic Interactions}
\subsection{The electroweak boson multispinor}
An examination of the form of the weak interactions according to the 
standard model, shows that the electroweak multiplet can be collected
into a bispinor
\be
{\cal W} \equiv -eQ\!\!\not\!A +g_WU\!\!\not\!W(1-i\gamma_5)/\sqrt{8}
      +g_W\!\!\not\!Z[T_3(1-i\gamma_5)-2q^2\sin^2\theta_W]/2\cos\theta_W,
\ee
where $U$ stands for the CKM matrix acting between `up' and `down' 
quarks, which obeys the unitarity property, $ U_i^j U_j^k = 
\delta_i^k;\quad U_j^k \equiv (U_k^j)^*,$
guaranteeing that the neutral weak interaction is flavour diagonal.
It is worth pointing out that the left-handed coupling of the weak 
bosons strictly applies to current quarks. There is reason to suspect 
that $V-A$ is not quite correct for constituent quarks, since it 
leads to an excessive $g_A/g_V$ ratio for the nucleon\footnote{This
is apparent on two counts: (i) the D/F ratio of 5/3 is rather larger 
than the experimental value $g_A/g_V\sim 1.25$ for nucleons, (ii) in 
supermultiplet theory the axial mesons represent the first orbital
excitation of the ground state mesons and have independent couplings 
to the quarks in principle. Really we should be writing the weak
couplings more accurately as the renormalized expression
$\bar{\psi}\!\not\!W(1-3i\gamma_5/4)\psi$, for constituent fields
$\psi$.}; but for the purposes of this paper we shall adhere to 
the picture of purely left-handed constituent quarks with the bonus 
of projection and simple Fierz reshuffling.

One of the primary elements of weak interactions is the pseudoscalar
meson decay constant for charged leptonic channels, which arises via
\be
 \langle 0|J_{\mu 5}^0|P^0\rangle \equiv if_Pp_\mu,\quad
 \langle 0|J_{\mu 5}^+|P^-\rangle \equiv i\sqrt{2}Uf_Pp_\mu;\quad
 J_{\mu 5} = \bar{\psi}i\gamma_\mu\gamma_5\psi.
\ee
We can think of this as a contact interaction between the weak bosons
and the pseudoscalar fields, in analogy to the vector dominance model:
\be
{\cal L} = f_P(W^0\!\cdot\partial P^0 +U\sqrt{2}W^+\!\cdot\partial P^-
           + iU^*\sqrt{2}W^-\!\cdot\partial P^+). 
\ee
If we {\em assume} that these $f$ have the {\em same} mass dependence
as in the vector case, on the basis of heavy quark theory \cite{HQS} 
or simply for uniformity, we would conclude that $f_P\simeq g_W
\sqrt{\lambda m_P}/4g$; this then leads to the predictions listed in
Table 2. At any rate, it is perfectly possible to reproduce the weak
boson interaction with the pseudoscalars through the multispinor
contact term
\bea
 {\cal L}_W&=& \frac{g_W}{4g} {\rm Tr}[{\cal W}\sqrt{\lambda m}\Phi]
        \supset \frac{g_W}{8g\sqrt{2}}{\rm Tr}[\!\not\!W
        (1-i\gamma_5)U(\!\not\!p+m_P)\gamma_5\phi_5]\nonumber\\
 & = & \frac{ig_W}{g\sqrt{8}}U W^\mu p_\mu\phi_5\sqrt{\lambda m_P}
     \equiv f_P\sqrt{2} UW.\partial\phi_5.
\eea
\begin{table}
\caption{Experimental values of pseudoscalar decay constants $f_P$,
 in MeV as determined from leptonic decay widths, compared with the 
 theoretical predictions, $f_P^2 \propto m_1+m_2$, where $m_i$
 are the constituent quark masses; note, for pseudoscalars we take
 $m_{ud} \simeq 200, m_s \simeq 300$, and of course $f_{\pi^0}=93$ 
 as our input.}
\vspace{0.2cm}
\begin{center}
\begin{tabular}{|l|c|c|c|c|c|} 
\hline
Pseudoscalar & $\pi^+$ & $K^+$ & $D^+$ & $D_s^+$ & $B^-$ \\ 
\hline
$f_P$ (expt.) & 131 & 160 & $<310$ & $240 \pm 40$ & ? \\
\hline
$f_P$ (theory) & 131 & 146 & 270 & 278 & 458 \\
\hline
\end{tabular}
\end{center}
\end{table}
This shows that we can regard the weak bosons as part of a bispinor
field; by coupling them to two pairs of quarks one arrives at the
flavour-changing nonleptonic weak interaction to order $G_F/\sqrt{2}
\equiv g_W^2/8m_W^2.$

\subsection{Nonleptonic amplitudes}
The standard treatment \cite{NL} is based on a current-current picture,
$$ {\cal H}_W = 4G_FU_{UD}U^*_{ud}[c_1(\bar{d}\gamma u)_L\cdot
                (\bar{U}\gamma D)_L+c_2(\bar{d}\gamma D)_L\cdot
                (\bar{U}\gamma u)_L]/\sqrt{2}, $$
where $c_i$ are Wilson coefficients. The next step is to invoke a
factorization hypothesis and evaluate this as sums of products,
$\langle h|J_L|h'\rangle\cdot\langle (P{\rm or}V)|J_L|0\rangle$, with
hadronic $h$ elements extracted from semileptonic decays; this despite
the fact that factorization does not work at all well for charmed meson
decays and does not properly apply to baryonic decays.

Using first principles and {\em permitting particle-mixing}, like
vector dominance and the old pole-model descriptions \cite{PM}, we will 
exhibit a supermultiplet scheme for calculating nonleptonic amplitudes 
more completely. Basically, we find that there are just three types of 
diagram which require evaluation to order $G_F$, as drawn in Figures 2,3. 
Fig 2 corresponds to a flavour-changing charge-conserving transition 
from one quark to another, Fig 3 to a neutral current transition from 
$d\bar{D}$ to $u\bar{U}$ and a charged transition from $d\bar{u}$ to 
$D\bar{U}$. Let us show how these may be estimated now (in Feynman gauge).

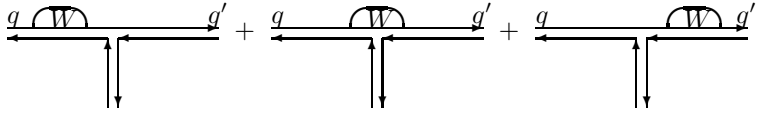
\begin{figure}
\begin{picture}(300,100)(-20,-30)
\put(20,2){$q$}
\put(96,2){$q'$}
\put(20,0){\vector(1,0){80}}
\put(58,-4){\vector(-1,0){38}}
\put(100,-4){\vector(-1,0){38}}
\put(58,-30){\vector(0,1){26}}
\put(62,-4){\vector(0,-1){26}}
\put(36,0){$W$}
\thicklines
\put(40,0){\oval(20,15)[tl]}
\put(40,0){\oval(20,15)[tr]}
\thinlines
\put(106,-4){+}
\put(120,2){$q$}
\put(196,2){$q'$}
\put(120,0){\vector(1,0){80}}
\put(158,-4){\vector(-1,0){38}}
\put(200,-4){\vector(-1,0){38}}
\put(158,-30){\vector(0,1){26}}
\put(162,-4){\vector(0,-1){26}}
\put(156,0){$W$}
\thicklines
\put(160,0){\oval(20,15)[tl]}
\put(160,0){\oval(20,15)[tr]}
\thinlines
\put(206,-4){+}
\put(220,2){$q$}
\put(296,2){$q'$}
\put(220,0){\vector(1,0){80}}
\put(258,-4){\vector(-1,0){38}}
\put(300,-4){\vector(-1,0){38}}
\put(258,-30){\vector(0,1){26}}
\put(262,-4){\vector(0,-1){26}}
\put(276,0){$W$}
\thicklines
\put(280,0){\oval(20,15)[tl]}
\put(280,0){\oval(20,15)[tr]}
\thinlines
\end{picture}
\caption{Weak quark-line flavour transition, including self-energy
and vertex contributions. \label{fig:Figure2}}
\end{figure}

\subsection{Quark-line transition}
The first part of Fig 2 includes a self-energy flavour change, eg $u 
\leftrightarrow c$, which produces the matrix element 
$\sum_q U^*_{uq}U_{cq}\Sigma_q(p)$ with
\be
 \Sigma_q(p)=\frac{ig_W^2}{2}\int\gamma_L\frac{1}{\not\!p\,+\not\!k-m_q}
             \gamma_L\frac{\bar{d}^4k}{k^2-m_W^2}  \simeq
        \left( \frac{g_Wm_q}{4\pi m_W} \right)^2 \frac{1}{2}\not\!p_L.
\ee
Note that a potential infinity in $\Sigma_q$ disappears by the unitarity
of $U$. Summing over $q$ and using known CKM elements, we obtain for $u
\leftrightarrow c$ say (in GeV), the total
$$\frac{\not\!p_LG_F}{\pi^2\sqrt{8}}[U_{ud}U_{cd}^*m_d^2 
 +U_{us}U_{cs}^*m_s^2+U_{ub}U_{cb}^*m_b^2] \simeq
 \frac{\not\!p_LG_F}{\pi^2\sqrt{8}}\times 
 .02 \sim 8\times 10^{-9}\not\!p_L.$$
Similarly, for the $d \leftrightarrow s$ transition (relevant for $K$
decays), one estimates the self-energy to be \cite{RDMDS}
$$\frac{\not\!p_LG_F}{\pi^2\sqrt{8}}[U_{us}U_{ud}^*m_u^2 
 +U_{cs}U_{cd}^*m_c^2+U_{ts}U_{td}^*m_t^2/4] \simeq
 - \frac{\not\!p_LG_F}{\pi^2\sqrt{8}}\times .14\sim
 -6\times 10^{-8}\not\!p_L.$$

The second part of Fig 2 is a vertex integral that reduces (for $d
\leftrightarrow s$ say) to $\sum_qU_{qd}U^*_{qs}I_q$, where
\be
 I_q=\frac{ig_W^2}{4}m_q\int\frac{(\not\!p_d\,+\not\!p_s-2\!\not\!k)_L
 \bar{d}^4k}{(k^2-m_q^2)^2(k^2-m_W^2)} \simeq \left(
  \frac{g_W}{8\pi m_W}\right)^2\frac{1}{2}m_q(\not\!p_s + \not\!p_d)_L. 
\ee
Observe that the internal quark mass weighting of $m_q$ is weaker than 
the $m_q^2$ factor in $\Sigma_q$, significantly so for the top quark.  
The third part is another self-energy at the other momentum leg $p'$.

Combining the three terms, we find that the combination is dominated by
the pole part associated with $\Sigma_q$, giving the total
\be
 \Gamma_W = \frac{1}{2}\sum_qU_{qs}U^*_{qd}\left(\frac{g_Wm_q}{8\pi m_W}
             \right)^2 \frac{m_s(1-i\gamma_5)-m_d(1+i\gamma_5)}{m_s-m_d},
\ee
which has to be traced out against the other quark line indices. For 
instance if we are studying purely pseudoscalar decay, the trace to be 
performed is
$$F{\rm Tr}[(\not\!p_2+m_2)\gamma_5\Gamma_W(\not\!p_1+m_1)\gamma_5
            (\not\!p_3+m_3)\gamma_5] .$$

\subsection{W-exchange diagrams}
The other two types of diagrams (see Fig 3) are essentially equal\footnote{
But if one takes $g_A/g_V\simeq 3/4$ for {\em constituent} quarks
their contributions become distinct.} by Fierz reshuffling and, apart 
from differing mass terms, give the typical pseudoscalar amplitude,
\bea
&\frac{Fg_W^2}{16m_W^2}U_{cs}U^*_{ud}{\rm Tr}[(\not\!p_c\!+\!m_c)\gamma_L
    (\not\!p_1\!+\!m_1)\gamma_5\gamma_L(\not\!p_u\!-\!m_u)
    (\not\!p_2\!+\!m_2)\gamma_5(\not\!p_3\!+\!m_3)\gamma_5]\nonumber\\
&=\frac{iFg_W^2}{4m_W^2}U_{cs}U^*_{ud}m_cm_u
\left[1-(\frac{m_c+m_u}{m_c-m_u})^2\right](m_2-m_3)(p_2\cdot p_3-m_2m_3).
\eea

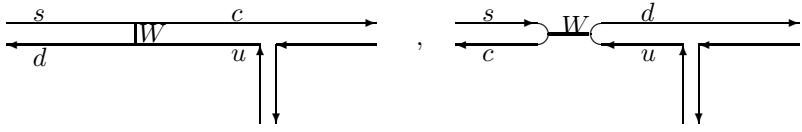
\begin{figure}
\begin{picture}(300,50)(0,-50)
\put(10,5){$s$}
\put(10,-12){$d$}
\put(0,4){\vector(1,0){140}}
\put(85,5){$c$}
\put(85,-10){$u$}
\put(96,-4){\vector(-1,0){96}}
\put(96,-34){\vector(0,1){30}}
\put(102,-4){\vector(0,-1){30}}
\put(140,-4){\vector(-1,0){38}}
\put(50,-3){$W$}
\thicklines
\put(49,4){\line(0,-1){8}}
\thinlines
\put(155,-4){,}
\put(170,4){\vector(1,0){30}}
\put(180,5){$s$}
\put(180,-10){$c$}
\put(240,5){$d$}
\put(240,-10){$u$}
\put(200,0){\oval(10,8)[r]}
\put(226,4){\vector(1,0){74}}
\put(256,-4){\vector(-1,0){30}}
\put(226,0){\oval(10,8)[l]}
\put(200,-4){\vector(-1,0){30}}
\put(256,-34){\vector(0,1){30}}
\put(262,-4){\vector(0,-1){30}}
\put(300,-4){\vector(-1,0){38}}
\put(210,0){$W$}
\thicklines
\put(205,0){\line(1,0){15}}
\thinlines
\end{picture}
\caption{$W$-exchange diagrams. The first is for a neutral quark 
combination and the second is for a charged combination.} 
\end{figure}

There are numerous processes to which we may apply these calculational 
methods and ideas, but space restrictions do not permit us to expose
any details. Work is currently in progress to evaluate reliably such weak
decay elements and we expect to present some of our results in the near 
future. At this stage all we can venture to say is that we estimate the 
self-energy transition is an order of magnitude (about 20 times) bigger 
than the $W$-exchange amplitude---automatically explaining the validity 
of the $\Delta I=1/2$ or `octet dominance' rule for strange particle 
decays. It is an auspicious sign.

\section*{Acknowledgements}
We thank the Australian Research Council for its support under grant 
number A69800907. Also, we are grateful to the CSSM for their hospitality 
during the QFT98 conference, when this manuscript was partially prepared.


\section*{References}
\begin{thebibliography}{99}
\bibitem{Revs}M. Wirbel, Prog. Nucl. Part. Phys. {\bf 21}, 333 (1988);
 J.G. Korner, D. Pirjol and M. Kramer, Prog. Part. Nucl. Phys. {\bf 33},
 757 (1990); J.D. Richman and P.R. Burchat, Rev. Mod. Phys. {\bf 67},
 893 (1996); T.E. Browder, K. Honschied and D. Pedrini, Ann. Rev. Nucl. 
 \& Part. Sc. {\bf 46}, 395 (1996); B.Stech in {\em b20 Symposium},
 Chicago, July 1997; M. Neubert in {\em Euroconference on QCD},
 Montpellier, July 1997.

\bibitem{RDDL1}R. Delbourgo and D. Liu, Phys. Rev. {\bf D53}, 6576 (1996).

\bibitem{RDDL2}R. Delbourgo and D. Liu, `Amplitude Analysis of Hadron 
 Decays', to appear in Phys. Rev. {\bf D}.


\bibitem{HQS}N. Isgur and M.B. Wise, \Journal{\PLB}{113}{1989};
 {\bf 237}, 527 (1990); M.B. Voloshin and M.A. Shifman, Yad. Fiz. 
 {\bf 47}, 801 (1988); H.D. Politzer and M.B. Wise, \Journal{\PLB}{206}
 {681}{1988}; E. Eichten and B. Hill, {\em ibid}. {\bf 234}, 511 (1990);
 H. Georgi, {\em ibid}. {\bf 240}, 447 (1990).

\bibitem{KT}J.G. Korner and G. Thompson, \Journal{\PLB}{264}{185}{1991}
 F. Hussain, J.G. Korner and G. Thompson, Ann. Phys. (NY), {\bf 206},
 334 (1991).

\bibitem{DSS}A. Salam, R. Delbourgo and J. Strathdee, Proc. Roy. Soc. 
 London Ser. {\bf A}284,146 (1965); M.A. Beg and A. Pais, Phys. Rev.
 Lett. {\bf 14}, 264 (1965); B. Sakita and K.C. Wali, Phys. Rev.
 {\bf 139}, B1355 (1965); R. Delbourgo {\em at al}, {\em The U(12) 
 Symmetry} (IAEA, Vienna, 1965).

\bibitem{NL}M. Bauer, B. Stech and M. Wirbel, Z. Phys. {\bf C34}, 103 
 (1987); J.G. Korner and M. Kramer, Z. Phys. {\bf C55}, 659 (1992); 
 B. Bajc, S. Fajfer and R.J. Oakes, Phys. Rev. {\bf D53}, 4957 (1996);
 M.A. Ivanov {\em et al}, Preprint hep-ph/9709325 and MZ-TH/97-21.

\bibitem{PM}P. Bedaque, A. Das and V.S. Mathur, Phys. Rev. {\bf D49},
 269 (1994); {\em ibid.} 1339 (1994) contain further references.

\bibitem{RDMDS}P. Pasqual and R. Tarrach, \Journal{\PLB}{87}{64}{1979};
 M.D. Scadron {\em ibid}. {\bf 95}, 123 (1980); B.H.J. McKellar and
 M.D. Scadron, Phys. Rev. {\bf D27}, 153 (1983); G. Eilam and M.D.
 Scadron, {\em ibid}. {\bf D31}, 2265 (1985); R. Delbourgo and M.D.
 Scadron, Nuovo Cim. Lett. {\bf 44}, 193 (1985); M.D. Scadron and
 D. Liu, `Top mass contribution to the $\Delta I$=1/2 rule', University
 of Tasmania preprint UTAS-PHYS-96-46.
 
\end{thebibliography}
\end{document}